\documentstyle[prl,eqsecnum,aps]{revtex}

\begin{document}
\draft
\title{The Dimer Model for $\kappa$--phase Organic Superconductors}

\author{Giovanni Visentini$^1$, Anna Painelli$^1$, Alberto Girlando$^1$
and Alessandro Fortunelli$^2$}
\address{$^1$Dipartimento di Chimica Generale ed Inorganica, Chimica Analitica
  e Chimica Fisica \\ Universit\`{a} di Parma, I--43100 Parma, Italy}

\address{$^2$Istituto di Chimica Quantistica ed Energetica Molecolare \\
 Consiglio Nazionale delle Ricerche, I--56126 Pisa, Italy}

\date{\today}
\maketitle
\begin{abstract}

We prove that the upper electronic bands of $\kappa$--phase 
BEDT--TTF salts are adequately modeled by an half--filled
tight--binding lattice with one site per cell.
The band parameters are derived from recent \emph{ab initio}
calculations, getting a very simple but extremely accurate
one--electron picture.
This picture allows us to solve the 
BCS gap equation adopting a 
real--space pairing potential.
Comparison of the calculated superconducting properties with
the experimental data points to isotropic $s_0$--pairing.
Residual many--body or phonon--mediated interactions offer a
plausible explanation of the large variety of physical properties
observed in $\kappa$--phase BEDT--TTF salts.

\end{abstract}

\narrowtext

Superconductivity (SC) in organic charge transfer (CT) 
salts has been discovered 
more than 15 years ago. 
At present, $\kappa$--phase BEDT--TTF (ET) salts
are the most promising organic superconductors (OSC)\cite{Williams}.  
OSC have similar features to 
cuprate superconductors:
they exhibit highly bidimensional structures, narrow electronic 
bands, low carrier densities and  possibly strong electronic correlations.
Superconducting properties are also similar in organics and cuprates, 
exhibiting singlet pairing \cite{Mayaffre},
very low  coherence lengths,   high magnetic
penetration depths and critical fields\cite{Williams}.
Perhaps the most characteristic feature of cuprate superconductors is 
the competition between SC and antiferromagnetism (AFM).
This competition shows up also in  OSC, as
demonstrated by the presence of antiferromagnetic spin fluctuations
in the metallic state of several compounds
\cite{Mayaffre}, or even by reentrant SC in the presence of
magnetic order. In this respect the different behavior of 
$\kappa$--ET$_2$Cu[N(CN)$_2$]X (ET-X) salts where X=Cl, Br 
is noteworthy: the two compounds are isostructural, but  at ambient pressure
the Br--compound is a superconductor with $T_c=11.6$~K,
whereas the Cl--compound is a Mott insulator with (possibly commensurate)
antiferromagnetic ordering\cite{Mayaffre}. On the other hand, even a very
small external pressure (270 bar), turns ET-Cl to the
superconducting ground state ($T_c=12.8$~K)\cite{Williams}. 
Experimental studies of normal and superconducting 
state properties in  OSC
are still scanty and often inconclusive, due to the extreme 
sensitivity of the material to pressure, radiation and sample
preparation. At the same time, the amount of theoretical 
work on OSC is scarce if compared with that developed for cuprates,
possibly due to the apparent complexity of the OSC structures.  
In the following we will show that, by a proper choice of the basis 
functions, the electronic structure of $\kappa$--phase ET salts can
be \emph{accurately} described in terms of a rectangular lattice with
nearest and next--nearest neighbor interactions. Similar one--band models 
are often adopted in the study of cuprates, however these models
lead to an oversimplified view of the complex cuprate 
band--structures\cite{Scalapino}.
OSC then constitute an unique opportunity to test SC theories within
a simple model, yet accurate enough to allow a \emph{quantitative}
comparison with experimental data.

The basic structural unit of $\kappa$-phases is a pair of ET
molecules. These pairs (dimers) arrange themselves nearly perpendicularly 
in a check-board
pattern to construct two dimensional cation layers intercalated by anion sheets.
As usual for organic CT salts, the relevant physics of ET compounds
is described in terms of electrons hopping among the frontier molecular
orbitals.
Due to their special structural motif, 
in $\kappa$-phases the intradimer interaction is larger than all
interdimer interactions. Therefore,
as already suggested by other authors \cite{Singleton,fuku},
the bonding and antibonding dimer orbitals are 
a convenient basis for tight--binding calculations.

The hopping integrals have been recently estimated for  ET-Br
by \emph{ab initio} calculations\cite{Fortu}.
The intradimer hopping $b_1$=0.272 eV
is more than two times larger than the interdimer integrals, $p$ =0.13 eV
and $q$ = 0.04 eV along $c+a$, and $b_2$ =0.085 eV along $c$ direction.
The dimer bonding and antibonding orbitals are split by an amount
$\approx 2b_1$, whereas interdimer hoppings give residual interactions  
$(p \pm q)/4$ and $b_2/2$. 
By neglecting the mixing between bonding and antibonding orbitals, 
the original four site problem reduces to two independent two site problems.
Since each dimer bears 3 electrons, the upper (conduction) band is half-filled.
Focusing on antibonding orbitals, each dimeric site has 2 nearest neighbors
along $c$ direction (interaction $b_2/2$), and 4 next-nearest neighbors
along $c \pm a$ directions (interaction $(p+q)/2$).
A fundamental advantage of the dimer model is that the dimer basis
fully exploits the local dimer symmetry, so that
the resulting tight--binding model is more symmetric 
than the underlying lattice.
The unit cell can therefore be reduced to contain a single site.
The analytical expression for the conduction band in the 
doubled Brillouin zone is:

\begin{displaymath} \epsilon({\bf k}) = b_2 \cos {k_z c} + 2(p+q)
                \cos {\frac{k_z c}{2} } \cos {\frac{k_x a}{2}}
\end{displaymath}

In fig.1 we report the energy dispersion curves  for the 
conduction and valence bands, the density of states (DOS) for the 
conduction band, and the Fermi surface (FS),
calculated in the dimer approximation, as compared with those 
obtained by solving
the tight--binding problem with four interacting site orbitals.
The main features of the complete
four orbital model are well reproduced in the dimer approximation,
namely, the  bandwidth, chemical potential, FS topology and DOS, and also
the location of the van Hove singularity (vHs).
The  bands in fig.1 compare well with EHT 
bands\cite{gei91},
justifying \emph{a posteriori} the neglect of inner orbitals.
Recently published LDA bands for $\kappa$--ET$_2$ Cu(NCS)$_2$ 
and ET--Br\cite{Ching} are also similar
apart from
an overall bandwidth narrowing.
Moreover the topology of the FS agrees
with the
experimental Shubnikov--de Haas (SdH) data on  
$\kappa$--ET$_2$Cu(NCS)$_2$\cite{Singleton} and on
$\kappa$--ET$_2$I$_3$\cite{Schweitzer95}.
Therefore the dimer model offers a good description of the $\kappa$--phase
band structure. We adopt this model to investigate SC.

Since in OSC the coherence length is  short (of
the order of few lattice spacings)\cite{Williams},
we choose a real space pairing potential 
analogous to that widely used  for 
cuprate superconductors\cite{Carbotte}:

\begin{equation}
V = -\sum_{i,j} g_{ij} a_{i\uparrow}^{\dag} a_{j\downarrow}^{\dag}
a_{j\downarrow} a_{i\uparrow} .
\label{pairing}
\end{equation} 

The sum is extended to nearest
and next--nearest neighbors (with coefficents $g_{NN}$ and 
$g_{NNN}$ respectively), and also includes $i=j$ (on--site) $g_{0}$ interaction, with 
either  positive (pairing) or negative (repulsion) coefficents. 
With the three parameters in Eq.~\ref{pairing} 
we explore all the possible symmetries of the gap function
$\Delta({\bf k})$ in the bidimensional point group \emph{pgg} of the 
organic layers of the crystal.
In particular, the BCS gap function for the singlet pairing can be
written as a linear combination of functions with different symmetries
\cite{Carbotte}:
\begin{math}
\Delta({\bf k}) = \sum_{i=1}^{4} \Delta_{i} \eta_{i}({\bf k}),
\end{math}
 where
\begin{math}
 \eta_{1}({\bf k})  =  \sin{\frac{k_{x}a}{2}} \sin{\frac{k_{z}c}{2}}
\end{math} 
transforms as the  $B_{g}$ irreducible representation, and
\begin{math} 
\eta_{2}({\bf k})  =  \cos{\frac{k_{x}a}{2}} \cos{\frac{k_{z}c}{2}},\ 
 \eta_{3}({\bf k})  =  1,\ \mathrm{and} \   
 \eta_{4}({\bf k})  =  \cos{k_{z}c}
\end{math}
as $A_{g}$.
Adopting the standard notation for cuprates\cite{Scalapino},
$\eta_{1}$ corresponds to $d_{xy}$ pairing and $\eta_{3}$ to $s_{0}$ 
isotropic pairing, while $\eta_{2}$ and
$\eta_{4}$ correspond to the extended or generalized $s^{\ast}$ 
pairing.
The $d_{xy}$ symmetry in the $\kappa$--phase
OSC corresponds  to  $d_{x^{2} - y^{2}}$  in the
cuprates, due to the different orientation of the
crystallographic axis with respect to the underlying lattice.
We observe that, at variance with cuprates, $d$-wave cannot mix with  
$s$-wave components.

Inserting the experimental $T_{c}$ = 11.6~K, relevant to ET-Br, 
into the gap equation,  
only two free pairing parameters survive, that we choose as
the ratios of $g_{0}$ and $g_{NN}$ over $g_{NNN}$.
We solve the BCS gap equation on a  1888$\times$1248 lattice 
in the $g_{0}$--$g_{NN}$ parameter space to obtain the phase diagram 
in fig.2. For several points in the phase diagram  
the calculated gap has been checked to
correspond to the absolute minimum of the BCS free energy.
At $T=T_c$ the phase diagram is very simple: the gap equation is linear 
so that the mixing of pair functions with different symmetry 
is strictly forbidden.  By lowering $T$ below $T_c$, the non-linearity
of the BCS equation allows the mixing of $s$ and $d$ pairing, and the mixing
region widens with decreasing  $T$.
However, the actual amount of mixing is always very small ($<10^{-3}$).
In fact, $T_c$ is rather low, so that the gap equation 
stays quasi-linear, and  the mixing of gap functions
with different symmetry is small\cite{Anderson}.

Having solved BCS gap equation, we  calculate the macroscopic properties
of the superconducting state using standard approaches\cite{BCS}.
Since in the  dispersion curves of the normal state 
the vHs  occurs at energies much higher than the
superconducting gap
(fig.1), the superconducting properties are largely dominated
by the states on the FS.
In particular, low energy macroscopic properties are strongly 
affected by the presence of nodes in the gap function at the FS.
On the other hand the maximum gap at the FS corresponds to a divergence in the 
superconducting DOS, and, as such, dominates frequency-dependent 
susceptibilities.
In fig.3 we report the minimum and maximum of $|\Delta({\bf k})|$ 
calculated at $T=0$. 
In the $d$--wave region, the minimum gap at the FS vanishes, i.e. nodes
are observed in the superconducting gap at the FS. In the  $s^{\ast}$--region
pseudonodes (not necessarily implying a change of
sign of the gap at the FS), are observed. 
At small $|g_{NN}|$ the $s$-wave gap is mainly isotropic and a sharp crossover 
is observed between $d$-wave and $s$-wave regions.
Due to the negligible mixing of $d$-- and $s$--pair  functions,
in the $d$-wave region
the maximum gap is fixed by the experimental $T_c$ to 2.09~$T_c$ (see Fig.3,
bottom panel).
In the $s$-region, due to the competition between isotropic and anisotropic
$s$-wave components the maximum gap at the FS shows a more complex structure.
We observe that for very small $|g_{NN}|$ the gap
is dominated by the 
isotropic component with the BCS value 1.76~$T_c$.

For the three representative  points marked in the phase diagram in  
fig.2, where
the gap is dominated by the isotropic, anisotropic $s$--wave
or $d$--wave contribution, respectively, we calculate 
the superconducting electronic specific heat,
magnetic penetration depth, and tunneling
bulk conductance.
The low--energy behavior of these quantities is 
exponential-like in the isotropic case, whereas it is power-law in the
anisotropic case, in agreement with standard results for conventional 
BCS\cite{BCS} and $d$-wave\cite{Maki} SC.
Experimental estimate of the specific heat\cite{Andraka} and
tunneling\cite{tunneling} are affected by large uncertainties,
so that the comparison is not conclusive. On the other hand,
experimental studies of the temperature dependence of the magnetic penetration
depth led to conflicting conclusions, suggesting either isotropic
pairing or anisotropic gapless SC\cite{magdepth}.
The superconducting
phonon self-energy, 
presents a singularity 
at a frequency corresponding to 
twice the maximum of the gap 
at the FS\cite{NJC}. By lowering $T$ below $T_c$,
phonons lying below (above) this frequency soften (harden).
The crossover frequency  strongly depends on 
the topology of the FS, so that the choice of the model 
for the electronic structure is now crucial.
The calculated crossover of the phonon self--energy does in fact
coincide with  $2\max |\Delta({\bf k})|$.
As shown in fig.3,  $\max |\Delta({\bf k})|$ 
reaches its minimum value for isotropic $s_{0}$ pairing, corresponding
to the BCS value $2\Delta_{0} = 3.53 T_{c} \approx 28$~cm~$^{-1}$.
The salient feature of a recent low--energy Raman scattering 
study\cite{io} is that below $T_{c}$ all relevant
bands harden with the greatest relative shift exhibited by the lowest
observed phonon at 27.4 cm$^{-1}$. This value corresponds to a
lower limit for the maximum of the gap at the FS.
On the basis of our model, this result can only be interpreted
assuming isotropic $s_{0}$ pairing.  Similar results hold for 
$\kappa$-(ET)$_2$Cu(NCS)$_2$, as can be inferred from recent neutron
scattering experiment\cite{neutron}.

The presented picture for SC in $\kappa$--phases is based on a single--particle
description of the electronic structure.
We have extracted the single--particle parameters from HF--SCF 
calculations\cite{Fortu}: the comparison with EHT\cite{gei91}
and LDA\cite{Ching} bands, as well as with experimental data on
$\kappa$--(ET)$_2$Cu(NCS)$_2$\cite{Singleton} and on
$\kappa$--(ET)$_2$I$_3$\cite{Schweitzer95} 
indicates that
the resulting picture is extremely accurate in reproducing
the topology of the FS.
On the other hand,
residual many--body effects  are important in determining
susceptibilities or dynamical properties of
OSC\cite{Singleton}.
For instance, the calculated electronic specific heat is quasi--linear with
$T$,  but the slope,
$\gamma \approx 7.5$ mJ/mole $\cdot$ K$^2$, is  
three times smaller than the experimental value, 
$22 \pm 3$ mJ/mole$\cdot$K$^2$ for ET--Br or
$24 \pm 3$ mJ/mole$\cdot$K$^2$ for $\kappa$--(ET)$_2$Cu(NCS)$_2$\cite{Andraka}.
Moreover, the effective masses inferred from SdH data via the
Lifshits--Kosevich formula for
$\kappa$--ET$_2$ Cu(NCS)$_2$\cite{Singleton} and 
$\kappa$--ET$_2$ I$_3$\cite{Schweitzer95} are about three times larger
than the single particle effective masses. 
It is important to stress that 
our picture for SC survives also in the presence of residual many-body
interactions,  in the hypothesis that these interactions
do not modify  the topology of the FS,
but only lead to 
an overall renormalization of single--particle parameters.
With the chosen
pairing potential, this
only implies an overall rescaling of pairing
parameters: for instance,
by rescaling our band parameters by a factor of 2,
the same phase diagram as in fig.2 is obtained, with 
pairing interactions reduced by a factor close to 2.

On the other hand,
the large variety of  physical properties characterizing
$\kappa$-phases suggests that residual interactions can also act to
modify the FS topology. 
In particular, ET--Cl is an antiferromagnetic
insulator, whereas
SdH oscillations are not observed in the (conducting) Br-analog at ambient
pressure. High pressure SdH oscillations in ET--Br point to a
distortion of the FS\cite{Kartsovnik}. 
In this connection,
the proposed dimer model is simple enough to offer clues about the
fine tuning of the physical properties of $\kappa$--phases
as induced by variations of the FS topology.
By looking at the band-structure in fig.1, one realizes that an 
\emph{extended}
vHs is present along the MZ direction,
at energy $-b_2$ with respect to the Fermi level.
Even a small reduction of $b_2$
would imply a large increase of the effective mass (due to the large effect
of the vHs), and a reduction of the $\alpha$ orbit, in agreement
with SdH measurement in ET-Br\cite{Kartsovnik}.
In the extreme limit of vanishing $b_2$
the vHS  would lie at the FS, and the FS itself would be a rectangle with
perfect nesting. This extreme picture seems expecially attractive to understand 
AFM in ET-Cl\cite{Mayaffre} as well as the
competition between SC and AFM
in ET-Cl and in ET-Br.
$b_2$ is a  small  parameter  and therefore it is subject to 
rather large relative variations in the different structures.
In Et-Cl and ET-Br salts the Cl and Br atoms are located very near to the $b_2$
dimer\cite{gei91}, possibly affecting the corresponding transfer integral.
\emph{Ab initio} calculations are in progress to test the effect
of the counterions and of
residual electron--electron interactions
on the 
$b_2$ hopping.

A different source of renormalization of the hopping parameters
involves electron--phonon interactions.
As it is evident from fig.1, there are
large regions in the $k$ space where the effective electronic masses are very 
large, even larger than the phonon masses.  In other words, since the
\emph{ab initio} $b_2$ value corresponds to 
685 cm$^{-1}$, so that most of the intramolecular 
phonons\cite{Bozio,Faulhaber}  are in the 
antiadiabatic limit with respect to $b_2$ hopping. Along this direction an 
antiadiabatic polaron narrowing\cite{Ciuchi} can become effective,
reducing $b_2$ hopping.
From  the available  coupling constants
and phonon frequencies\cite{Bozio,Faulhaber}, 
the polaron narrowing factor is estimated about 3.

In summary, we have shown that 
an extremely simple model -- a tight-binding lattice with one
site per cell -- offers an accurate description of the band
structure of $\kappa$--phase ET salts.
Using this model we were able to get a
solution of the full, non--linear BCS gap equation for OSC.
At variance with cuprates, our simple one--band model 
can be derived from first--principle calculations, 
and therefore allows us to get a
significant comparison with experiment.
Our work supports isotropic $s_0$ SC.
Finally, the possible role of counterions as well as of phonon--mediated
interactions is invoked to justify the large variability of physical
properties of $\kappa$--phase ET salts.

\noindent We thank  R. Bozio and M. Acquarone for valuable discussions.
Work supported by the Italian Consiglio Nazionale delle Ricerche (C.N.R.)
and by the Ministero dell'Universit\`{a} e della Ricerca Scientifica e
Tecnologica (M.U.R.S.T.).

\pagebreak

\noindent {\bf Figure captions}\par
\bigskip
\noindent \emph{Figure 1.} The tight--binding highest occupied
bands (left panel), DOS (right panel), and FS (inset)  calculated for
$\kappa$-(ET)$_2$Cu[N(CN)$_2$]Br
in the dimer approximation (solid line),
and for the full four site problem (dotted line).\par
\bigskip
\noindent \emph{Figure 2.} The three--dimensional phase diagram for
SC. Dashed lines are the isotherms for temperatures below $T_c$.
Points lying between two isotherms correspond to mixed $s$-- and
$d$--wave states.
The circle, diamond, and star indicate representative points for 
isotropic
($g_{NNN} = 4.96 \cdot
10^{-2}$ eV), anisotropic $s$--wave 
($g_{NNN} = 4.76 \cdot 10^{-2}$ eV), and
$d_{xy}$--wave 
($g_{NNN} = 7.24 \cdot 10^{-2}$ eV), respectively.\par
\bigskip
\noindent \emph{Figure 3.} The minimum (top panel) and
maximum (bottom) of $|\Delta({\bf k})|$ on the Fermi Surface at
$T = 0$, as a function of the relative pairing parameters strengths (see
text).

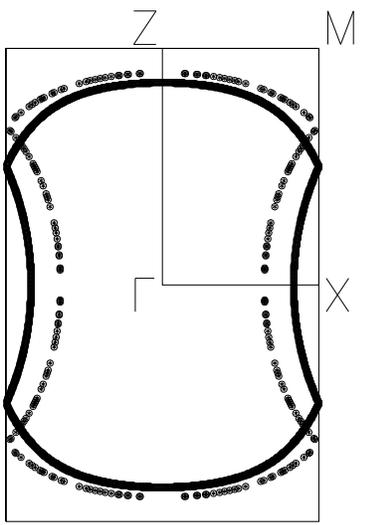
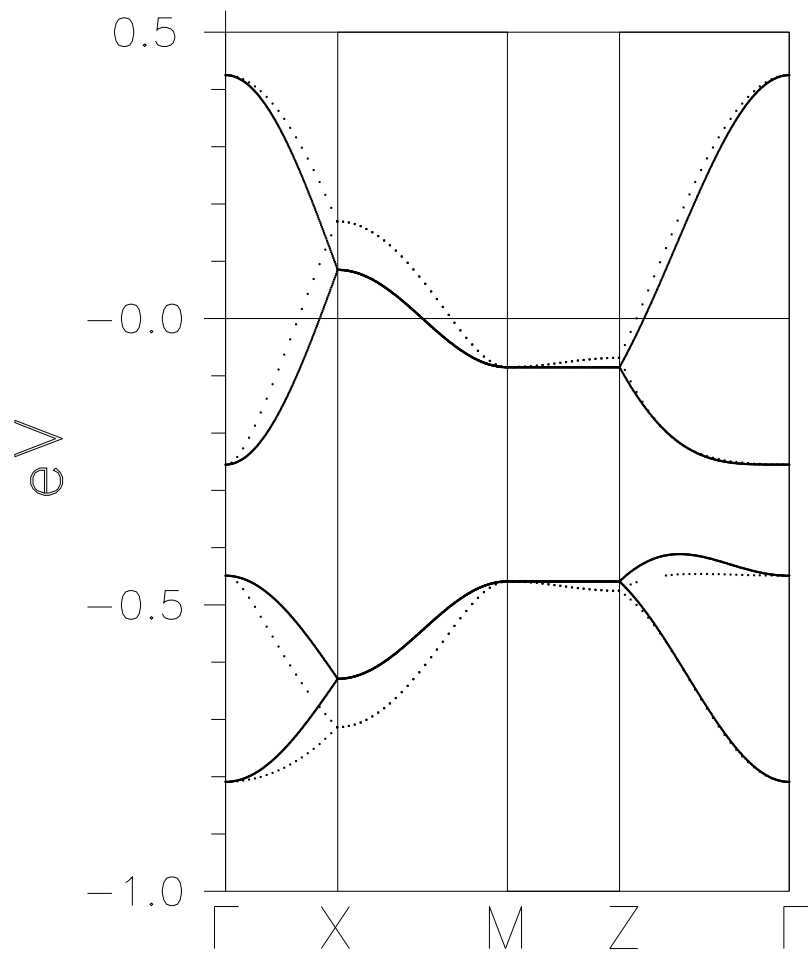
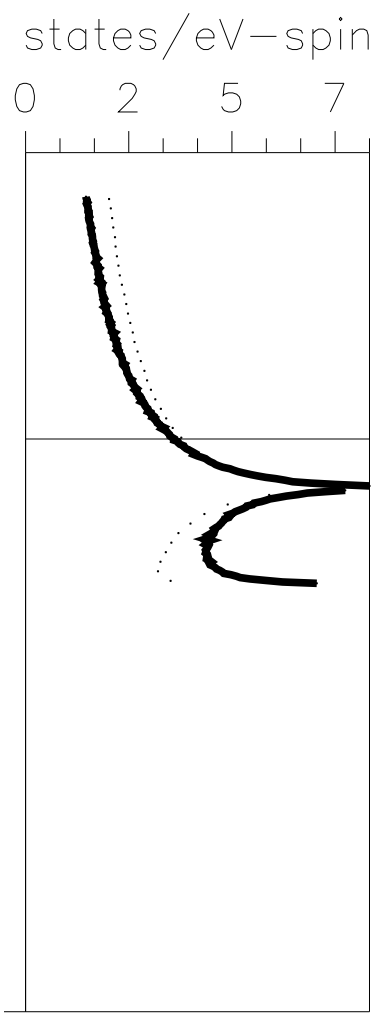
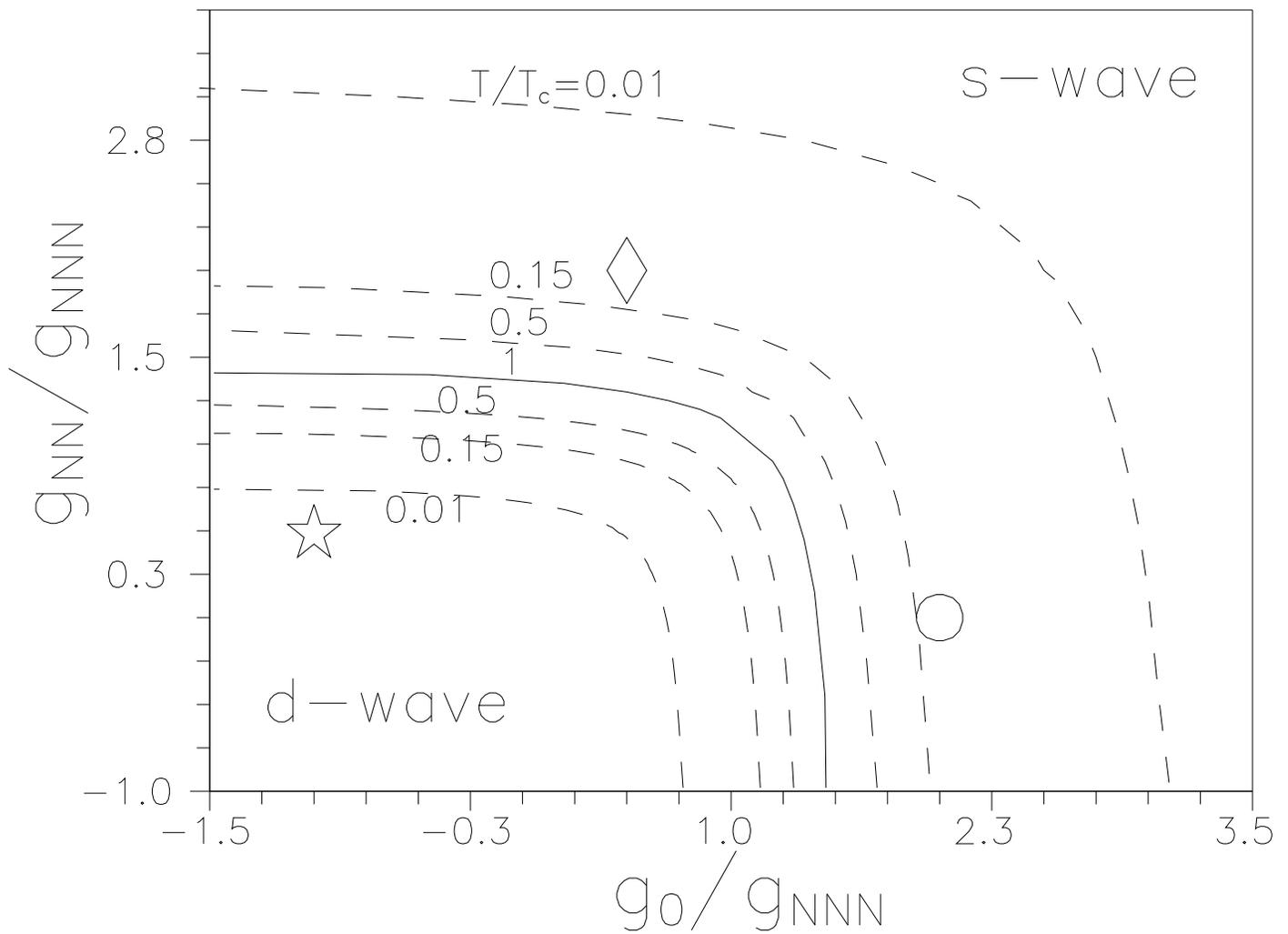
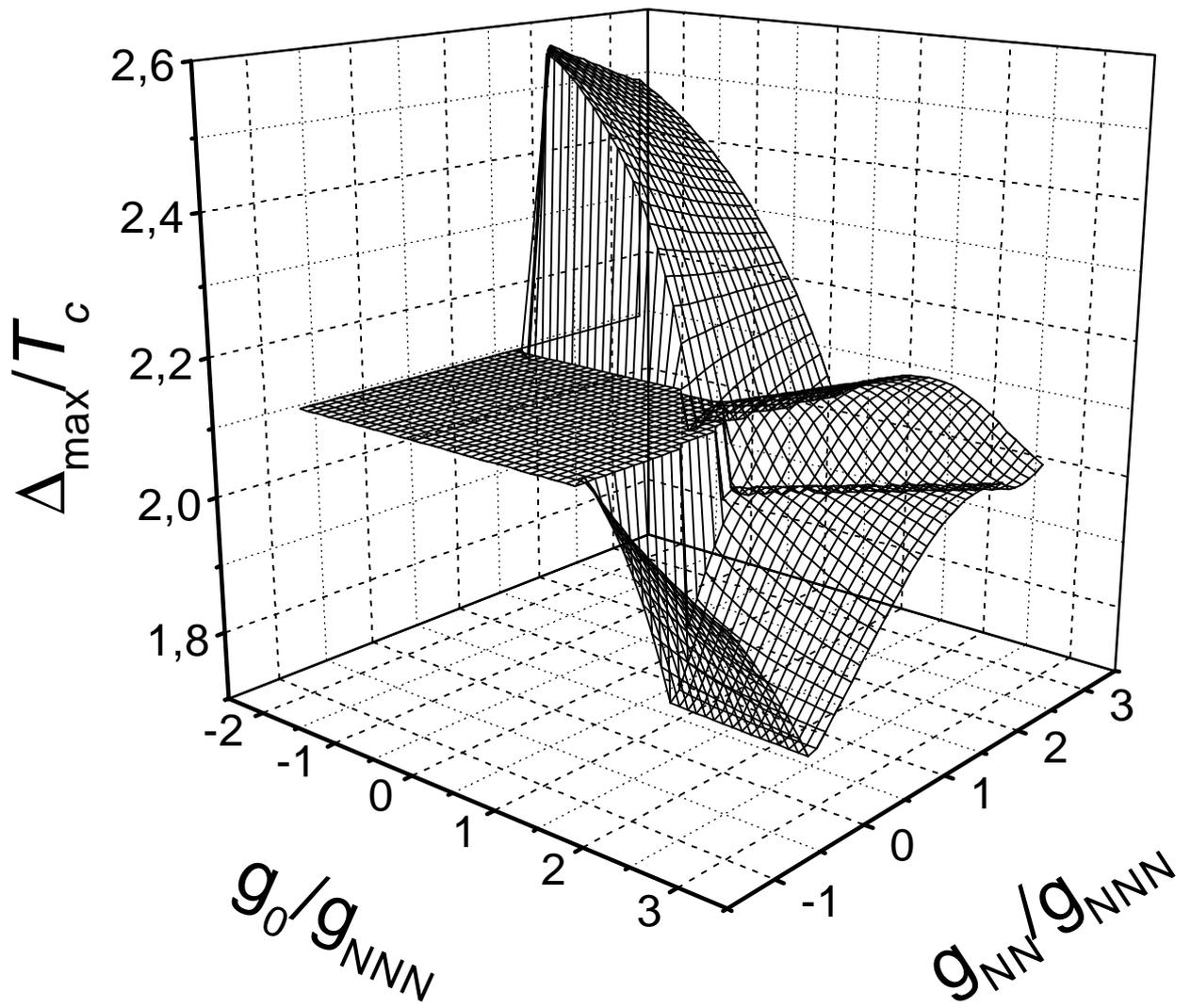
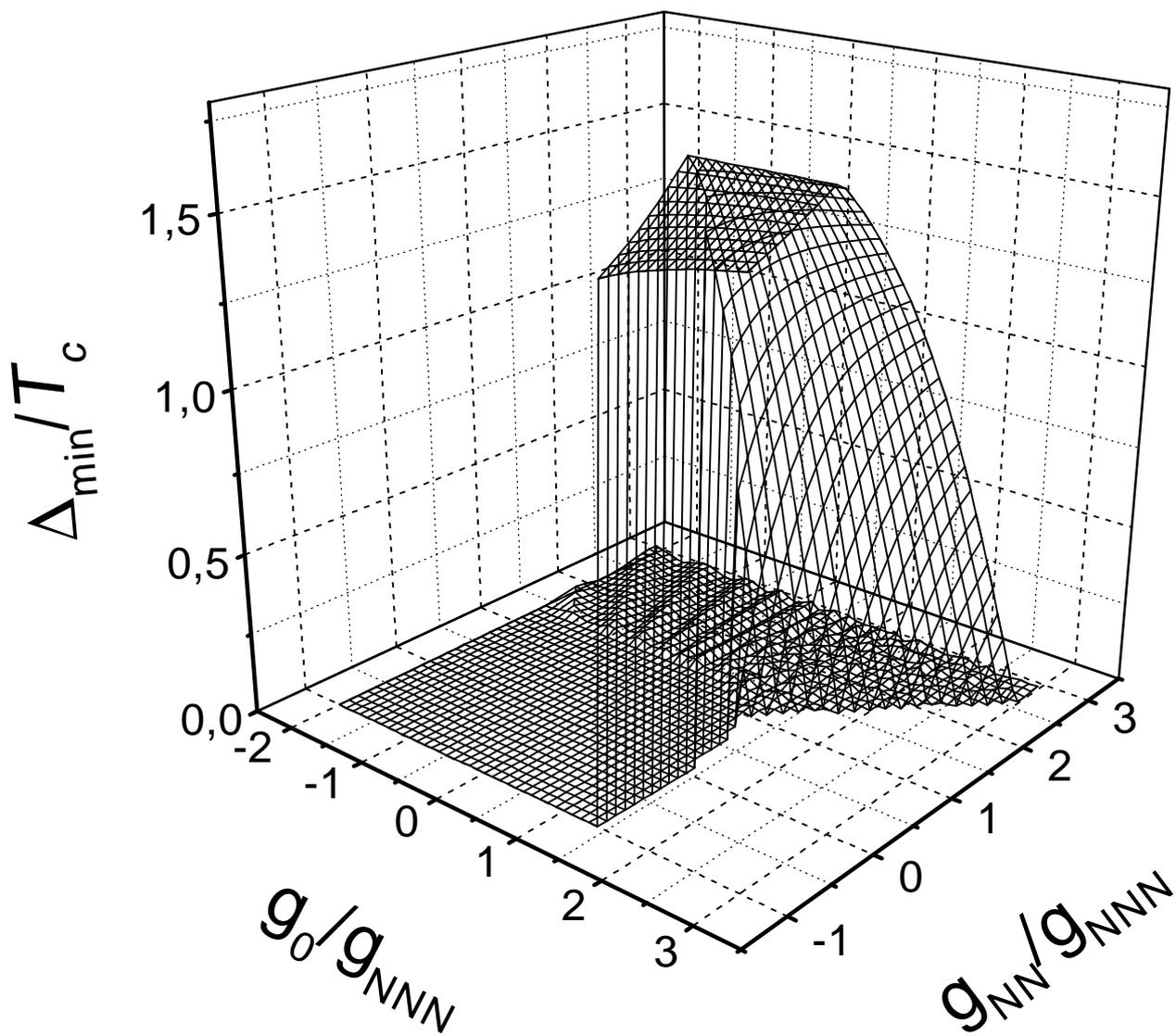
\end{document}